\documentclass[a4paper,11pt]{article}
\usepackage{jheppub} 
\usepackage{lineno}
\usepackage[utf8]{inputenc}
\usepackage[english]{babel}
\usepackage[T1]{fontenc}

\usepackage{graphicx}%
\usepackage{multirow}%
\usepackage{amsmath,amssymb,amsfonts}%
\usepackage{amsthm}%
\usepackage{mathrsfs}%
\usepackage[title]{appendix}%
\usepackage{physics}
\usepackage{amssymb}
\usepackage{xcolor}
\usepackage{tikz}
\usetikzlibrary{quantikz2}
\usepackage{caption}
\usepackage{subcaption}

\newtheorem{theorem}{Theorem}

\title{Out of the box approach to Black hole Information paradox}
\author[a]{Kiran Adhikari}
\affiliation[a]{Emmy Noether Group for Theoretical Quantum Systems Design, Technical University of Munich, Germany}
\emailAdd{kiran.adhikari@tum.de}
\author[b]{}
\affiliation[b]{}
\abstract{put abstract here...}
\begin{document}

\abstract{

Suppose a black hole forms from a pure quantum state $\ket{\psi}$. The black hole information loss paradox arises from semiclassical arguments suggesting that, even in a closed system, the process of black hole formation and evaporation evolves a pure state into a mixed state. Resolution to the paradox typically demands violation of quantum mechanics or relativity in domains where they should hold. Instead, I propose that in a complete theory of quantum gravity, any region $\mathcal{U}$ that could collapse into a black hole should already be described by a mixed state, thus bypassing the paradox entirely. To that end, I present a model in which the universe is in a quantum error-corrected state, such that any local black hole appears mixed and encodes no information locally.}


\maketitle

\section{Introduction}

Because the fundamental laws are reversible, the conservation of information is sometimes regarded as the zeroth law of physics. In principle, one could reconstruct a lump of coal purely from the radiation it emits. However, Hawking showed that by applying the accepted laws of quantum field theory to accepted laws of classical curved spacetime, one reaches a paradoxical conclusion that the initial state of the black hole cannot be reconstructed out of its radiation \cite{Hawking:1975vcx}. This is dubbed the infamous information loss problem, which somehow keeps making people go around in circles indefinitely \cite{Mathur:2009hf, unruh2017information, polchinski2017black, Witten:2024upt}.

To get a clear hold of the problem, we can compare three different entropy curves for a black hole of mass \( M \). The first is the Hawking Curve, which represents the von Neumann entropy of Hawking radiation as computed in \cite{Hawking:1975vcx}. It starts at zero and increases to order \( GM^2 \) as the black hole evaporates completely. The second is the Bekenstein-Hawking curve, which represents the thermodynamic entropy of the black hole itself \cite{PhysRevD.7.2333, Hawking:1975vcx}. It starts at order \( GM^2 \) and decreases to zero as the black hole evaporates. The final is the Page curve that represents the von Neumann entropy of Hawking radiation under the assumption that black hole evolution is unitary \cite{Page:1993wv, Hayden_2007}. If the black hole forms from a pure state, the Page curve initially increases, reaching approximately \( \frac{GM^2}{2} \) at the Page time (\( t_{\text{Page}} \)), before decreasing back to zero as the black hole evaporates. If the black hole instead forms from a mixed state with entropy \( GM^2 \), the Page curve follows the Hawking curve exactly since no late-time radiation is needed to purify the earlier emissions.

If a black hole starts in a pure state, the difference between the Page curve and the Hawking curve at \( t_{\text{Page}} \) requires an order-one change, possibly through mechanisms like wormholes, firewalls, or quantum extremal islands \cite{Almheiri:2012rt, Maldacena:2013xja,Penington:2019npb, Almheiri:2019qdq, Almheiri:2019psf}. This order-one change is also a reason why any solutions to the paradox seem so dramatic. 

However, purity in quantum mechanics is a subjective and observer-dependent quantity—what appears pure to one observer may be mixed with another. Consider a system \( AB \)  in a maximally entangled EPR state, then both \( A \) and \( B \) are maximally mixed with respect to both observers. When \( A \) measures his qubit, entanglement is lost, making \( B \)'s state pure and known from \( A \)'s view. However, until \( A \)'s signal reaches \( B \), \( B \) still sees its state as mixed. This puts us in an awkward situation: If purity is subjective, how can it cause an objective physical change for a black hole?

To further explore this, I now present the following thought experiment where \( A \) and \( B \) start as maximally entangled Bell pairs and $B$ collapses into a black hole. Suppose $A$ and $B$ are separated by a distance much greater than the lifetime of \( B \)'s black hole (\(\text{distance}_{AB} \gg 2t_{\text{Page}}\)). Around the Page time \( t_{\text{Page}} \), \( A \) measures his qubits, destroying entanglement with \( B \) and making \( B \)'s state pure from \( A \)'s perspective. If \( A \) hadn’t been measured, \( B \)’s Page curve would have followed a mixed-state trajectory. After the measurement, it should instead follow a pure-state trajectory. Since $A$'s measurement occurs outside $B$'s light cone, no causal signal can reach 
$B$ in time to “know” whether $B$ should develop islands, wormholes, or other new physics needed to change the Page curve in such a drastic way. 

One possible conclusion might be that restoring the Page curve may not be the correct approach to solving information loss. Instead, we might have to reconsider the very notion of purity in quantum gravity. 

\section{Main Idea}

 Suppose a black hole of mass $M$ is formed from a pure state $\ket{\psi}$. The black hole information paradox is a conclusion that these four postulates cannot be mutually consistent.
\begin{enumerate}
\item Postulate 1: Semiclassical gravity is valid in the low curvature region, including near the black hole horizon.
    \item  Postulate 2: As seen from the outside, a black hole can be described as a system with $ S = \frac{A}{4GN}$ degrees of freedom, where $S$ is the entropy and $A$ is the area of its event horizon. 
\item Postulate 3: The final state of the Hawking radiation, following the complete blackhole evaporation, is in a pure state. 
\item Postulate 4: The equivalence principle is valid. A freely falling observer crossing the event horizon should experience no special local effects and should not be able to tell when they pass the event horizon based on local physics alone.
\end{enumerate}

Resolution to black hole information paradox usually revolves around relaxing one or more of these postulates. For example, fuzzball and firewall proposals relax postulate 1 \cite{mathur2005fuzzball, Almheiri:2012rt, MersiniHoughton:2014cta}, while remnants and final burst relax postulate 2 \cite{giddings1992black, bekenstein1994entropy, ong2024case}, and information loss \cite{unruh2017information} challenges postulate 3. Proposals such as ER = EPR \cite{Maldacena:2013xja}, blackhole complementarity \cite{susskind1993stretched} and other holographic variants contradict \cite{maldacena2023ads,Maldacena:1997re} postulate 4 by requiring the horizon to somehow encode the infalling object's information. However, Quantum mechanics says that it is impossible for the horizon to gain information about the infalling state without disturbing it \cite{fuchs1998information}. Such disturbance can be detected by the infalling observer, leading to a violation of the equivalence principle. Therefore, none of these solutions are fully satisfactory, as they either demand violations of quantum mechanics or general relativity in domains where they should hold or introduce significant non-locality or acausality.

Here, I argue that the assumption that a black hole can form from a pure state should be reconsidered—an approach I titled an out-of-the-box solution. In a full theory of quantum gravity, the very concept of a pure state might not even be well-defined in the same way as in standard quantum mechanics. In fact, we already expect from quantum field theory that any finite region of space is necessarily in a mixed state due to entanglement with its surroundings. If there is no pure state to begin with, then we might bypass the paradox entirely. I thus propose the following: 
\begin{quote}
 In a theory of quantum gravity, any region $\mathcal{U}$ that could collapse into a black hole should be described by a mixed state.   
\end{quote}

This proposal contrasts with the Unruh-Wald approach \cite{unruh2017information}, where black holes are modeled as open systems, allowing a pure initial state to evolve into a mixed one and thus permitting information loss.

\section{Principle of information conservation}
\label{sec:principleOfInformationConservation}
Suppose a system $A$ evolves into a system $B$. One could define information to be conserved if the quantum state describing $A$, denoted as $\rho_A$, can be reconstructed from the state of $B$, $\rho_B$, via some recovery operation $\mathcal{R}$, such that
\begin{equation}
\mathcal{R}(\rho_B) = \rho_A.
\end{equation}
This definition has been discussed in the context of black holes, particularly in the seminal works of Page and Hayden-Preskill \cite{Page:1993wv, Hayden_2007}.

Other notions of information conservation exist in quantum gravity, such as those arising from holographic dualities. Specifically, certain information in $A$ may be accessible from $B$ if the partition functions of the two systems are equal $Z_A = Z_B$ \cite{witten1998anti}, or if any arbitrary correlation function in $A$,
\[
\langle O_1 O_2 \dots O_n \rangle_A,
\]
can be determined from certain correlation functions in $B$ via a recovery operation $\mathcal{F}$,
\[
\langle O_1 O_2 \dots O_n \rangle_A = \mathcal{F}(\langle \phi_1 \phi_2 \dots \phi_m \rangle_B),
\]
where $O_i$ are operators in $A$, $\phi_i$ are operators in $B$, and $\mathcal{F}$ represents the recovery operation or mapping between the two systems' correlation functions \cite{Maldacena:1997re, Bousso:2002ju}.

However, in this work, I adopt the former definition of information conservation, as it accounts for both coarse and fine-grained information. For this, I will rely on one of the central results in entanglement theory, quantum error correction, a technique where initial seed information is distributed across a larger multipartite entangled system such that even after erasing a certain number of subsystems, initial seed information can still be recovered. Bell's theorem and quantum error correction are probably the non-trivial results in entanglement theory \cite{NielsenChuang2000}. Therefore, it is no wonder that quantum error correction, though a seemingly boring engineering topic,  plays such a fundamental role for us.

 There exist certain criteria known as Knill-Laflamme conditions \cite{Knill:1997} for quantum error correction. These conditions can also be formulated in terms of information-theoretic quantities such as entropy and mutual information \cite{Grassl_2022, NielsenChuang2000}. This reformulation is especially valuable in the context of quantum gravity, where the notion of quantum states may be ill-defined, yet entropy is still expected to remain well-defined \cite{Witten:2021unn}.

Let the seed information $S$, described by a quantum state $\rho_S$ in a Hilbert space $\mathcal{H}_S$ of dimension $K$, be purified by a reference system $R$ with Hilbert space $\mathcal{H}_R$. The resulting purification is denoted by $  |S R\rangle \in \mathcal{H}_S \otimes \mathcal{H}_R$. If the seed information is encoded in qubits, the number of qubits required would be \( \log_2 K \). To proceed forward, I will now define two information theoretic quantities: von Neumann entropy and mutual information. The von Neumann entropy of the quantum system $X$ with density matrix $\rho_X$ is given by:
\begin{equation}
S(X) = -\operatorname{Tr} (\rho_X \log \rho_X) = -\sum_{j=1}^m \lambda_j \log \lambda_j,
\end{equation}
where $\lambda_1, \dots, \lambda_m$ are the eigenvalues of $\rho_X$. The quantum mutual information between two quantum systems $X$ and $Y$ is defined as:
\begin{equation}
I(X:Y) = S(X) + S(Y) - S(XY),
\end{equation}
where $S(X)$, $S(Y)$, and $S(XY)$ are the von Neumann entropies of $X$, $Y$, and the joint system $XY$, respectively.

We denote by $\ket{RX_1 \dots X_n}$ the state of the quantum system $RX_1 \dots X_n$ after applying a completely positive isometric encoding map $V$ to $S$ and identity to $R$. The $n$ quantum systems $X_1, \dots , X_n$, each of level $q$, is partitioned into two arbitrary disjoint blocks $X_I$ and $X_J$, of size $|I|=n-d+1$ and $|J|=d-1$, respectively. A quantum erasure error correcting code of distance $d$ allows the reconstruction of the seed information $\rho_S$ even after erasing block $X_J$. This is possible if the following two information-theoretic requirements are satisfied \cite{r9}:
\begin{itemize}
    \item Recoverability: Recovery of seed information: $I(R:X_I) = I(R:S)$
    \item Secrecy: No information by erasures: $I(R:X_J)=0$
\end{itemize}
 If the seed information $S$ is pure, entropic arguments provide a bound \cite{NielsenChuang2000}, known as quantum singleton bound, for the number of such erasures: 
\begin{equation}
    \log_q K\leq n-2d+2.
\end{equation}
The singleton bound implies $|J| = d - 1 < \frac{n}{2}$ for $K > 1$, indicating that two disjoint blocks cannot independently reconstruct the seed information—an argument consistent with the no-cloning theorem \cite{wootters1982single}. 
For the case when the seed information $S$ starts in a mixed state, then the $X_1, \dots, X_n$ would also be in a mixed state. Suppose one needs to add $s$ more extra systems to purify it. Then, the bound for the number of such erasures satisfies:
\begin{equation}
\label{eq:singletonBoundEntropic}
  d-1 < \frac{n+s}{2}.
\end{equation}
The result of Page \cite{Page1999ThePC} and Hayden-Preskill \cite{Hayden_2007} can also be retrieved from such singleton bounds \ref{sec:RampSecret}. Indeed, Page results belong to the case when $S$ is pure, while  Hayden-Preskill belongs to the case when $S$ is mixed.

Of course, perfect recovery might not be possible, especially in the context of naturally occurring scrambling systems such as black holes and cosmology, and we might have to be satisfied with approximation. Fortunately, given the recoverability requirement $I(R:X_I) = I(R:S)$,  Uhlmann’s theorem \cite{Uhlmann1976} guarantees the existence of a recovery map $\mathcal{R}$, such that upon acting on $\rho_{X_I}$ gives out the approximate $\rho_S$, $\mathcal{R} \left( \rho_{X_I}\right) \approx \rho_S $. Similarly, the secrecy requirement $I(R:X_J)=0$ implies that the system $\rho_{X_J}$ is decoupled from the reference system $R$: $\rho_{X_J R} = \rho_{X_J} \otimes \rho_{R} $ meaning the system $R$ is uncorrelated from $X_J$ and is therefore statistically independent of any measurement on $X_J$. Interestingly, this implies that any subsystem of size up to $d-1$ is maximally mixed:
\begin{equation}
\label{eq:maximallyMixed}
    \rho_{X_J} = \frac{\mathbb{I}_{X_J}}{\text{dim}(X_J)}
\end{equation}
where $\text{dim}(X_J)$ is the Hilbert space dimension of $X_J$.

So far, our discussion has been fairly abstract. However, two key physical properties of quantum error-correcting codes are particularly relevant for us. First, isometric encoding maps are generally not unitary and cannot be directly implemented in a closed system. Instead, they usually require the introduction of extra degrees of freedoms $\ket{0}_A,$ such that the encoding process follows:  $V|\psi\rangle = U\left(|\psi\rangle \otimes |0\rangle_A\right).$ Second, the encoding map must generate entanglement between the seed information and ancillas. This entanglement is the key ingredient that protects quantum information from errors. Interestingly, many naturally occurring chaotic systems, including random unitaries, are pretty good sources of entanglement. Therefore, these chaotic systems, often referred to as scramblers in literature \cite{Sekino_2008}, naturally satisfy the conditions for approximate quantum error correction. The appendix \ref{sec:scrambling} and \ref{sec:RampSecret} outlines the relationships between different notions of information scrambling and the criteria for quantum error correction.

\section{Cosmological information conservation}

I will now present a framework in which the seed information of the universe, described by a quantum state $\rho_S$, at or before the Big Bang remains preserved throughout the evolution of the universe, including during the formation and evaporation of black holes. This is facilitated by modeling the wave function of the universe as a quantum error-corrected state. The following two features are particularly relevant for this: 
\begin{enumerate}
    \item  Expansion as Ancilla factory: The quantum error correction is possible because of the isometric map from a smaller Hilbert space to a larger one. The physical implementation of the isometric map requires extra quantum degrees of freedom, known as ancillas. Fortunately, the expanding universe serves as an excellent ancilla factory. Indeed, authors from \cite{cotler2023isometric,CotlerStrominger2022} proposed that time evolution is isometric for quantum gravity in expanding cosmologies with concrete examples such as matter effective field theory in rigid curved spacetimes and de Sitter quantum gravity, namely, Jackiw-Teitelboim with a positive cosmological constant. \cite{bao2017quantum} also presented a quantum circuit model for the expanding cosmology, where initially unentangled ancilla becomes entangled as expansion continues.
    \item Big Bang as a scrambler: While we lack a complete theory of the Big Bang, we expect it to be a highly complex and chaotic system. A trick, due to Wigner \cite{Wigner1957}, is to model such complex systems using random matrices. Since random unitaries are known to be effective encoders of quantum information, we model the Big Bang as a scrambler that encodes seed information into a highly entangled state. This perspective is further supported by the Susskind–Sekino conjecture \cite{Sekino_2008}, which proposes that black holes are the fastest scramblers in nature, and it's reasonable to extend this concept to the Big Bang itself.
\end{enumerate}

These two features—the ancilla-like degrees of freedom emerging from the expanding universe and the Big Bang acting as a scrambler—together suggest that the ancillae become highly entangled with the seed information as time evolves. As a result, the quantum state of the universe takes the form of a quantum error-corrected state resilient to erasure errors. 

\subsection{Black holes as erasures and the information paradox}

Suppose the seed information, $\rho_{S}$, is encoded into $n$ quantum systems $X_1, \dots, X_n$, which can be purified by adding other $s$ quantum systems. 
 If the black hole $B$ is composed of $x$ such quantum systems, then they can be modeled by erasure errors if the size $x$ is less than $\frac{n+s}{2}$. As discussed in section \ref{sec:principleOfInformationConservation}, such a black hole would be in a maximally mixed state described by a density matrix:
 \begin{equation}
     \rho_{B} = \frac{\mathbb{I}_{B}}{\text{dim}(B)}
 \end{equation}
 where,  $\text{dim}(B)$ is the Hilbert space dimension of the black hole $B$. This framework offers a potential resolution to the black hole information paradox without requiring any departure from quantum mechanics or general relativity. Since the black hole starts in a mixed state, the entanglement entropy of the black hole remains unchanged before and after evaporation. Information is thus never localized within the black hole itself. Instead, the seed information is distributed nonlocally across the entire universe. When a black hole evaporates, no information is lost as the global state remains unchanged. 

Let the total size of the independent black holes formed from the $n$ quantum subsystems be denoted by $X_B$. Such black holes can be tolerated as long as $X_B$ satisfies the bound:  
\begin{equation}
X_B < \frac{n + s}{2},
\end{equation}
where $s$ is the extra quantum systems needed to purify $n$ quantum systems $X_1, \dots, X_n$.  If the seed information starts in a pure state, $s = 0$, and the condition reduces to $X_B < \frac{n}{2}$, similar in principle to the situation of Page \cite{Page_1993}. This implies that the universe can tolerate erasures (i.e., black holes) affecting up to half of its total subsystems without losing the encoded seed information. 

\begin{figure}
    \centering
   \begin{subfigure}[b]{0.8\textwidth}
   \centering
         \includegraphics[width=\textwidth]{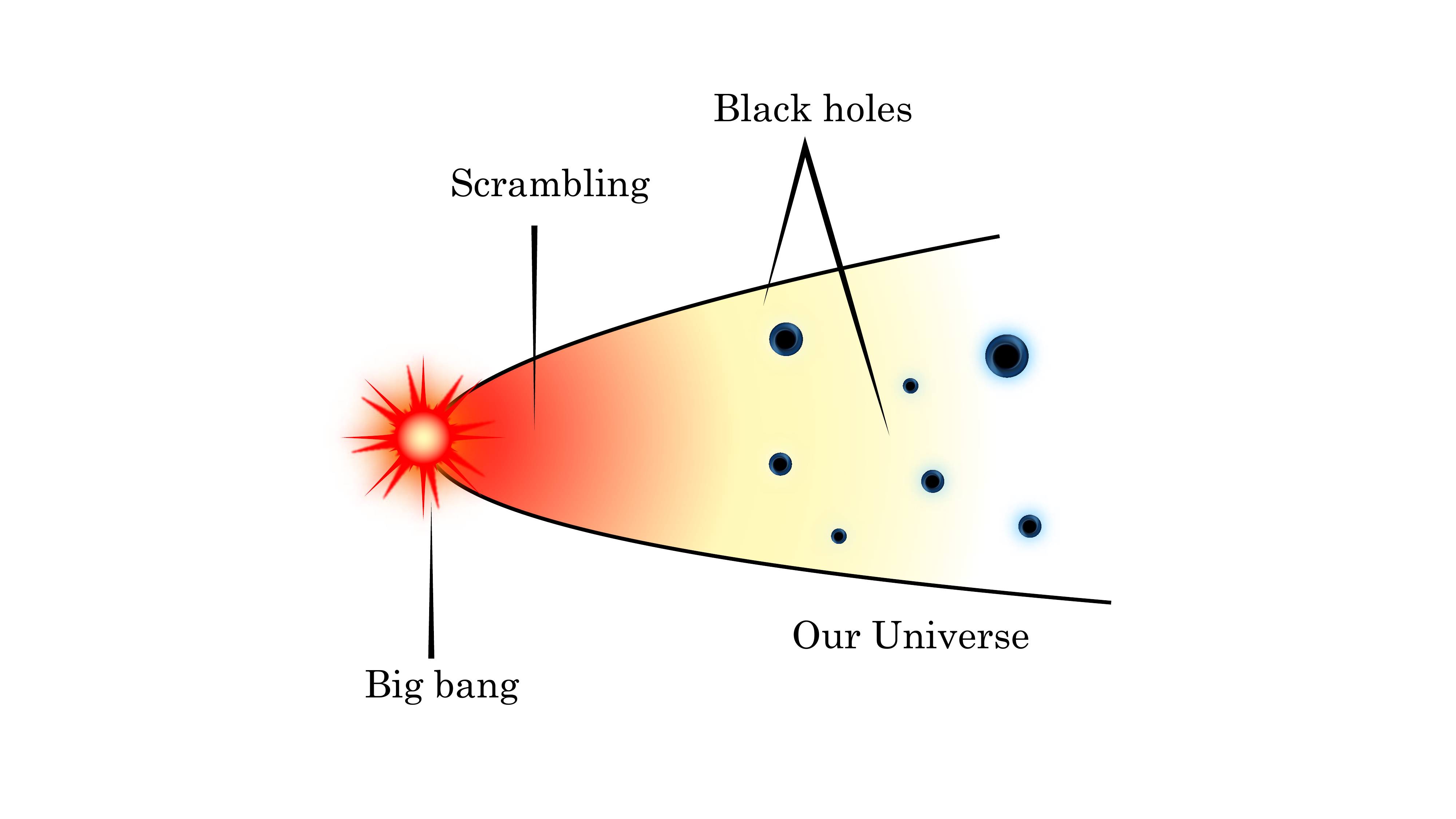}
         \caption{Suppose the seed information during the Big Bang began in a pure state, which becomes scrambled with ancillas as the universe expands. The universe in, then, in a quantum erasure-corrected state. Black holes are modeled as erasures and thus appear as mixed states and carry no information.}
         \label{fig:universeMirror}
   \end{subfigure}
   \hfill
    \begin{subfigure}[b]{0.8\textwidth}
   \centering
         \includegraphics[width=\textwidth]{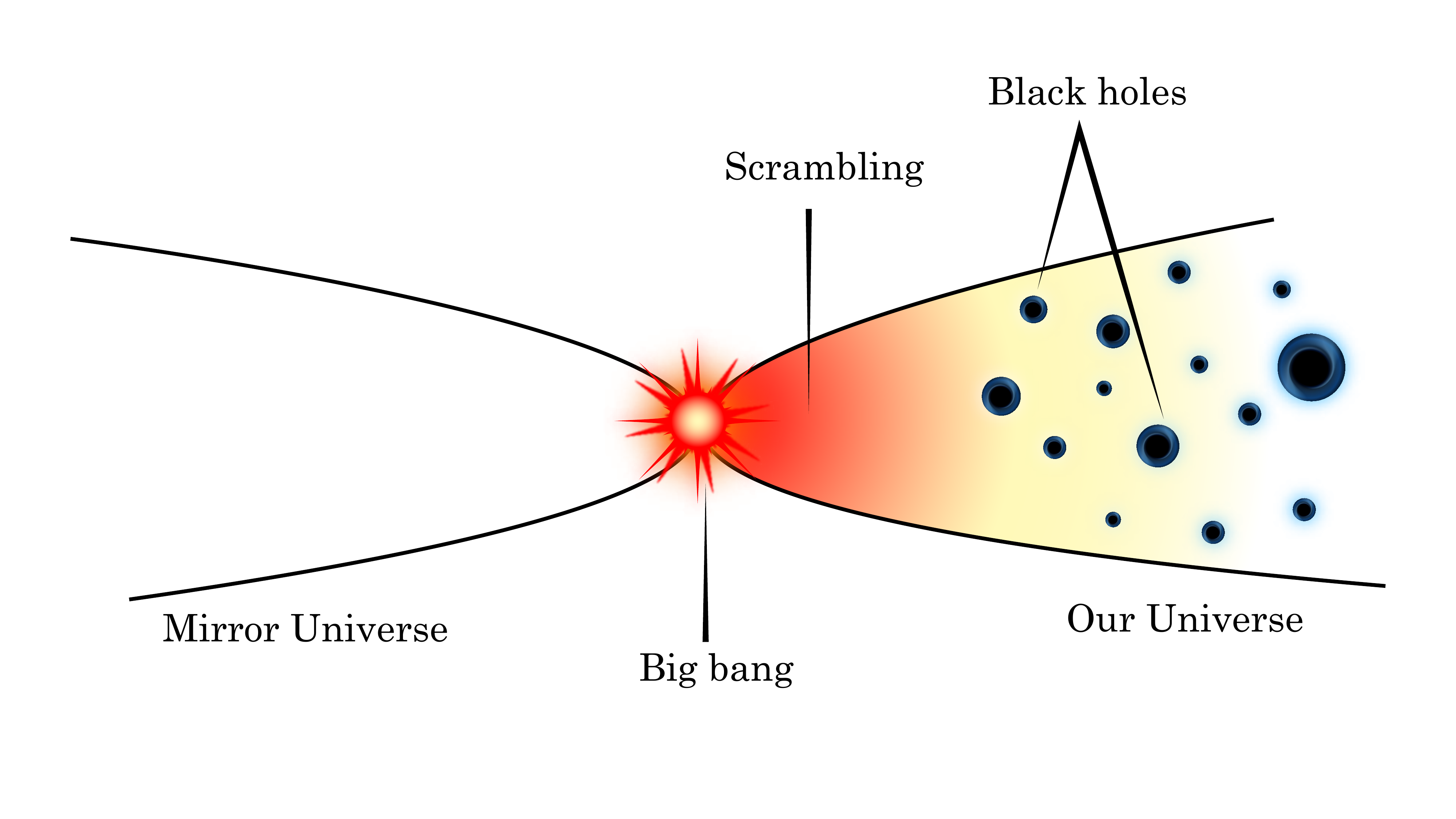}
         \caption{If, instead, the seed information starts in a mixed state, possibly entangled with the mirror universe, more black holes can be tolerated.}
         \label{fig:quantumCircuitModel}
   \end{subfigure}
   \hfill
    \begin{subfigure}[b]{0.8\textwidth}
   \centering
         \includegraphics[width=\textwidth]{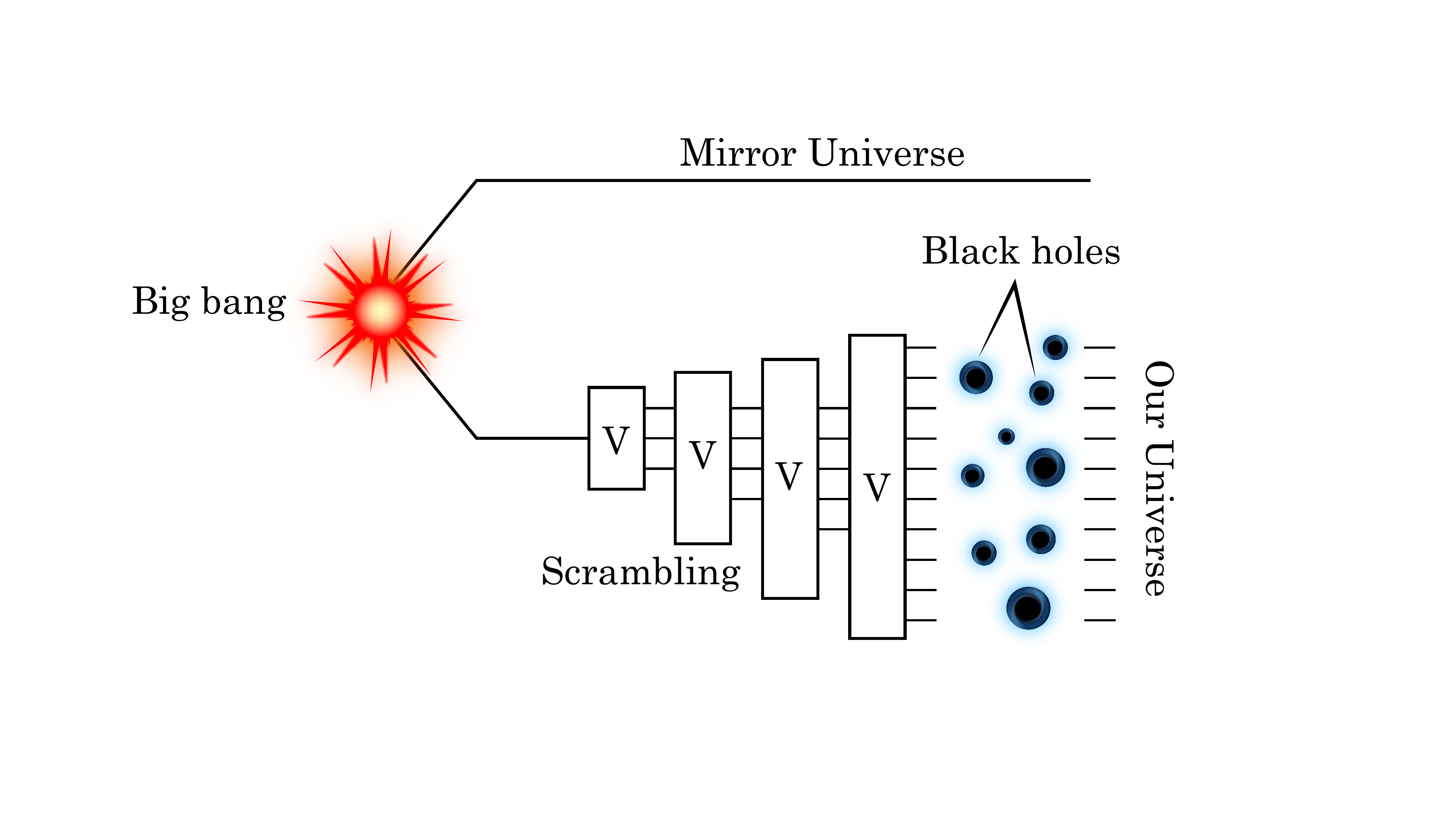}
         \caption{Quantum Circuit Model}
         \label{fig:quantumCircuitModel3}
   \end{subfigure}
   \caption{Cartoon illustrating Cosmological Information Conservation}
    \label{fig:universeModel}
\end{figure}

A more interesting case happens when the seed information starts in a maximally mixed state, and $s \approx n$. Then, the number of black holes that can be tolerated is roughly:
\begin{equation}
    X_B \approx \frac{n+n}{2} = n.
\end{equation}
Even if almost the entire universe, excluding the arbitrary subsystem of the size equal to the size of the seed information $\rho_S$, collapses into a black hole, the seed information from the Big Bang remains conserved. Such a scenario might be possible if the early universe emerged in a highly entangled state with a ``mirror universe,'' forming a global pure state, as illustrated in figure \ref{fig:universeModel},  described by a thermofield double (TFD)---like structure:
\begin{equation}
    |\Psi\rangle_{\text{tot}} = \sum_i \sqrt{p_i} \, |i\rangle_{\text{universe}} \otimes |i\rangle_{\text{mirror}},
\end{equation}
where $\{p_i\}$ forms a probability distribution determined by the initial entanglement structure. By tracing out the mirror sector, the observable universe appears as a thermal mixed state:
\begin{equation}
    \rho_{\text{universe}} = \sum_i p_i \, |i\rangle\langle i|.
\end{equation}
If $p_i = \frac{1}{d}, \forall i$, where $d$ is the rank of entanglement or equivalently infinite temperature case of TFD state, $\beta =0$, the reduced state of the universe would be in a maximally mixed state and $X_B \approx n$. 

Indeed, the metric of the past universe is simply given by $g_{\mu \nu}=a^2(\tau) \eta_{\mu \nu}$, where $\eta_{\mu \nu}$ is the flat Minkowski metric, and the scale factor $a(\tau)$ is just proportional to the conformal time $\tau$. This metric exhibits additional isometry under time-reversal symmetry  $\tau \rightarrow-\tau$, and gives a new analytically extended mirror universe. The mirror universe is also sometimes referred to as an antiuniverse under the $CPT$ symmetric universe hypothesis \cite{Boyle:2018tzc}. Relevant for us is that the mirror universe would be in a highly entangled state with our universe, thus error corrects the mess our black hole creates. Such an entangled origin may also arise from a gravitational vacuum via a Euclidean path integral with two boundaries, as in the Hartle-Hawking no-boundary proposal \cite{HartleHawking1983} or in the formation of the baby universe pairs as Reeh-Schlieder theorem \cite{ReehSchlieder1961} suggests that the vacuum is maximally entangled at all scales, including Big Bang. 

\section{Discussions/Conclusions}
As usual, attempts to solve the information paradox raise deep philosophical questions on locality and purity, and the approach in this article is no exception. 

First, let's consider the locality. Suppose Alice occupies a spacetime region \( \mathcal{U} \), and let \( \mathcal{U}^C \) represent the rest of the universe. In principle, Alice can be converted into a black hole. If we were to believe Hawking's original semiclassical calculation, Alice's original information wouldn't be recoverable from the radiation itself. However, information cannot be lost and must then be located in $\mathcal{U}^C $. The no-cloning theorem of quantum mechanics suggests that it would be impossible to recover the same quantum information from two disjoint regions.

This raises a question: was Alice ever located in the region $\mathcal{U}$? The answer appears to be no—not in the fundamental sense. This is the kind of non-locality and emergent space-time, the cosmological information conservation suggests. Note that this locality is not the same as  Einstein's locality, which is rooted in causality via finite signal propagation. Here, the non-locality emerges not from faster-than-light signaling but from the deeper structure of quantum information and entanglement itself. It would be interesting to explore the connection with similar ideas, such as the principle of holography of information, which suggests that cosmological correlators in an arbitrarily small region are sufficient to fully specify the global state \cite{chakraborty2023holography}.

Second, the principle of cosmological information conservation implies that any finite, local region $\mathcal{U}$ is described by a mixed state. This raises a natural question: how, then, do we reconcile this with the common laboratory assumption that a quantum system can be prepared in a pure state $\ket{\psi}$? The answer seems to be that, at the fundamental level, in terms of quantum field theory (QFT) or quantum gravity, such questions would have highly non-trivial answers \cite{Witten:2021jzq, sorce2024notes}. In fact, a classic result due to Araki \cite{Araki1968, Araki1974} is that von Neumann algebra for a local region $\mathcal{U}$ in the case of QFT in the flat Minkowski spacetime is of Type $III$. While a type $I$ algebra that describes the usual laboratory quantum mechanics has a notion of pure states, density matrices, or von Neumann entropies, Type $III$ algebras lack these because the trace is not defined for them.

To clarify, we define a trace as a complex-valued linear function $a \rightarrow \operatorname{Tr}$ a that satisfies $\operatorname{Tr} a b=\operatorname{Tr} b a$ and a positivity condition $\operatorname{Tr} a^{\dagger} a>0$ for all $a \neq 0$. For either Type $I$ or Type $II$, the trace can be defined, and correspondingly density matrices and other information theoretic quantities such as $S(\rho) = \operatorname{Tr}  (\rho \log \rho)$. Recently, \cite{Witten:2021unn} showed in the context of quantum gravity, the crossed-product construction can transform Type $III$ algebras to Type $II$. These results suggest that the Type $II$ is particularly relevant in gravity, as entropies can be defined even in the absence of well-defined quantum mechanical microstates as in a local laboratory. Since the cosmological information conservation principle is formulated in terms of such information-theoretic quantities, this suggests that while a well-defined notion of purity and microstates may not exist in a local laboratory setting, global information conservation remains possible.

Our proposal seriously considers the possibility that spacetime may be emergent from quantum entanglement. It would be interesting to explore the connection with other holographic approaches \cite{Harlow2015, almheiri2015bulk,Harlow2018} and other emergent gravity theories\cite{jacobson1995thermodynamics, jacobson2016entanglement, verlinde2011origin, PhysRevD.97.086003} that aim to reconstruct spacetime geometry from more fundamental, often information-theoretic, principles.

At the moment, our proposal appears more like a scaffold for a much deeper theory -- much like Maxwell’s early mechanical models of the electromagnetic field. A compelling direction for future work is to develop this framework into a more concrete formalism and to explore its implications for phenomenological questions, such as the primordial black holes, the nature of dark matter, and dark energy.

\begin{appendices}

\section{A short intro to Quantum Information Theory}
In this section, we provide a brief review of the concepts from quantum information theory relevant to this paper. For a more comprehensive introduction to these concepts, interested readers can refer to the literature \cite{qcn,r6}. Let us consider a quantum system $X$ consisting of $m$ degrees of freedom modeled by the algebra of $m \times m$ matrices over the complex numbers, denoted by $\mathcal{M}_m$. The state of $X$ is described by its density matrix $\rho_X \in \mathcal{M}_m$. In classical information theory, Shannon entropy provides a measure of uncertainty of a random variable, whereas von Neumann entropy generalizes it to the quantum case. The von Neumann entropy of $X$ with density matrix $\rho_X$ is given by:
\begin{equation}
S(X) = -\operatorname{Tr} (\rho_X \log \rho_X) = -\sum_{j=1}^m \lambda_j \log \lambda_j,
\end{equation}
where $\lambda_1, \dots, \lambda_m$ are the eigenvalues of $\rho_X$. Using quantum entropy, one can introduce a quantum generalization of classical mutual information, called quantum mutual information, which gives a measure of the correlation between different quantum systems.  Let us consider two quantum systems, $X$ and $Y$, with density matrices $\rho_X$ and $\rho_Y$, respectively. The quantum mutual information between $X$ and $Y$ is defined as:
\begin{equation}
I(X:Y) = S(X) + S(Y) - S(XY),
\end{equation}
where $S(X)$, $S(Y)$, and $S(XY)$ are the von Neumann entropies of $X$, $Y$, and the joint system $XY$, respectively. In addition to the von Neumann entropy, there exists a quantum version of the classical R\'enyi entropy, called the quantum R\'enyi entropy. For a given quantum system $X$ with density matrix $\rho_X$, the R\'enyi entropy of order $\alpha$ is defined as:
\begin{equation}
S^\alpha(X) = \frac{1}{1-\alpha} \log \text{Tr}[\rho_X^\alpha],
\end{equation}
where $\alpha \geq 0$ and $\alpha \neq 1$. For integer values of $\alpha$, the R\'enyi entropy only involves integer powers and traces of the density matrices, making it easier to compute for physically relevant situations. Moreover, in the limit $\alpha \rightarrow 1$, the R\'enyi entropy reduces to the von Neumann entropy:
\begin{equation}
\lim_{\alpha \rightarrow 1} S^\alpha(X) = S(X).
\end{equation}
The R\'enyi entropy is upper bounded by $\log d$, where $d$ is the dimension of the Hilbert space of the system, and this upper bound is achieved if and only if $\rho_X$ is a maximally mixed state, i.e., $\rho_X = \frac{\mathbf{1}}{d}$. We can also define the R\'enyi version of mutual information between two quantum systems $X$ and $Y$ as:
\begin{equation}
I^\alpha(X:Y) = S^\alpha(X) + S^\alpha(Y) - S^\alpha(XY).
\end{equation}

Frequently, there is a need to compare two different quantum states. Such a comparison can aid in, for instance, evaluating the output of an ideal quantum protocol against a noisy one. One possible method is to employ the operator trace norm, also referred to as the $L_1$ norm. The $L_1$ norm for any operator $M$ is defined as follows:

\begin{equation}
||M||_1 = \text{Tr} \sqrt{M^\dagger M}.
\end{equation}
This norm can be used to calculate the distance between two density matrices $\rho$ and $\sigma$, which is represented by $||\rho - \sigma||_1$. The trace norm distance is motivated physically because for $||\rho - \sigma ||_1 < \epsilon$, $\text{Tr}(\Pi (\rho - \sigma)) < \epsilon$ for any projection operator $\Pi$. This implies that the probability outcome of any experiment between two density matrices $\rho$ and $\sigma$ differs by at most $\epsilon$. Once the distance between two density matrices is established, the quantum mutual information can be employed to place an upper bound on the correlation between two distinct quantum systems. This bound is also referred to as quantum Pinsker's inequality \cite{r6}. The normalized trace distance between $\rho(XY)$ and $\rho(X) \otimes \rho(Y)$ can be defined as $\Delta = \frac{1}{2} ||\rho(XY) - \rho(X) \otimes \rho(Y)||_1$. Then quantum Pinsker's inequality implies \cite{r6}:
\begin{equation}
    \frac{2}{\ln 2} \Delta^2 \leq I(X:Y)
\end{equation}

When the mutual information $I(X:Y)$ is small, it indicates little correlation between the two quantum systems $X$ and $Y$. Thus the joint state $\rho(XY)$ can be approximated by the tensor product of the marginal states $\rho(X)$ and $\rho(Y)$. Another measure of distance between two quantum states is the fidelity, which is defined as $F(\rho(X), \rho(Y)) = (\text{Tr} \sqrt{\rho(X)\rho(Y)})^2$, where $\sqrt{M}$ denotes the unique positive square root of a positive semidefinite matrix $M$. The trace distance can be used to provide upper and lower bounds on the fidelity through the Fuchs-van de Graaf inequalities \cite{r6}, which are given by 
\begin{equation}
    1 - \sqrt{  F(\rho (X),\rho (Y) )} \leq \Delta  \leq \sqrt{1 - F(\rho (X),\rho (Y) ) }
\end{equation} where $\Delta$ is the normalized trace distance. The fidelity is bounded below by $0$ and above by $1$. Two density matrices $\rho(X)$ and $\rho(Y)$ are said to be $\mu$-distinguishable if $F(\rho(X), \rho(Y)) \leq 1-\mu$, and $\nu$-indistinguishable if $F(\rho(X), \rho(Y)) \geq 1-\nu$.


Another extremely important concept in quantum information theory is the decoupling inequality, which has wide-ranging applications such as quantum error correction, transmission of quantum information, protocols such as state merging, coherent state merging, thermodynamics, many-body physics, and even black hole information theory \cite{Hayden_2007, decoupling, Grassl_2022, Sekino_2008}. Suppose a composite system $XY$ is in a joint state $\rho(XY)$. The subsystem $X$ is decoupled from $Y$ if: $\rho(XY) = \rho(X) \otimes \rho(Y)$. This means that the system $X$ is uncorrelated from $Y$ and is therefore statistically independent of any measurement on $Y$. One common way to analyze this setup is to start with the pure state system $RA$, Alice's message $A$ purified with reference $R$. After this, we encode Alice's message $A$ with unitary encoding and allow it to interact with the environment, which gives us the final output $BE$. $B$ here can represent Bob's received message. If the encoding was done such that $R$ and $E$ decouple, then this implies that there exists a decoding protocol such that just having system $B$, Bob can recover Alice's message $A$. This works even for the case when $R$ and $E$ are approximately decoupled, $\rho(RE) \stackrel{\epsilon}{\approx} \rho(R) \otimes \rho(E)$, that is,
\begin{equation}
F(\rho(RE),\rho (R) \otimes \rho (E)) \geq 1 - \epsilon \text{ for some } \epsilon \geq 0
\end{equation}
Uhlmann's theorem \cite{r6} guarantees that there exists a decoder such that having system $B$, one can approximately reconstruct Alice's message $\rho (A') \stackrel{\epsilon}{\approx} \rho (A)$, that is:
\begin{equation}
F(\rho (A'), \rho (A)) \geq 1 - \epsilon \text{ for some } \epsilon \geq 0
\end{equation}
We will later make use of the decoupling theorem to define the notion of quantum information scrambling.

We will also frequently make use of properties of Haar integrals. An integral over the unitary group $U$ ($2^n \times 2^n$) of the matrix function $f(U)$ with respect to the Haar measure can be represented as:
\begin{equation}
I = \int dU f(U)
\end{equation}
The defining property of the Haar measure is left- (respectively, right-) invariance with respect to shifts via multiplication. If $V$ is a fixed unitary, then:
\begin{equation}
\int dU f(UV) = \int d(U'V^\dagger) f(U') = \int dU'f(U')
\end{equation}


In order to properly analyze the setting, we divide the input state as $AB$ and output state $CD$. Furthermore, the unitary $U$ is taken to be random $2^n \times 2^n$ matrix taken from Haar ensemble. The Haar
average lets us consider expectations over a number of unitary matrices and is non-zero
only when the number of $U$ equals the number of $U^\dagger$.
For example with the case of two $U$s and two $U^\dagger$s, we get:
\begin{equation}
\label{eq:haarintegral}
\begin{aligned}
       \int dU U_{i_1j_1}U_{i_2j_2}U^*_{i'_1j'_1}U^*_{i'_2j'_2} = \frac{1}{2^{2n}-1}&  (\delta_{i_1 i'_1}\delta_{i_2 i'_2}\delta_{j_1 j'_1}\delta_{j_2 j'_2} + \delta_{i_1 i'_2}\delta_{i_2 i'_1}\delta_{j_1 j'_2}\delta_{j_2 j'_1})\\
       &- \frac{1}{2^n(2^{2n}-1)} (\delta_{i_1 i'_1}\delta_{i_2 i'_2}\delta_{j_1 j'_2}\delta_{j_2 j'_1}  + \delta_{i_1 i'_2}\delta_{i_2 i'_1}\delta_{j_1 j'_1}\delta_{j_2 j'_2})
\end{aligned}
\end{equation}
Let $d= d_Ad_B = 2^a 2^b= d_Cd_D = 2^c2^d = 2^{a+b} = 2^{c+d} = 2^n$, then $d = 2^n$.

This formula will let us compute the average over the trace of the square of the density matrix $\rho_{AC}$:
\begin{equation}
    \int dU \text{Tr}(\rho^2_{AC}) = \frac{1}{2^{2n}} \int dU U_{klmo}U^*_{k'lm'o}U_{k'l'm'o'} U^*_{kl'mo'}
\end{equation}
where, $k = 1,...,2^a$ are $A$ indices, $l = 1,......,2^b$ are $B$ indices, $m = 1,.....,2^c$ are $C$ indices, and $o = 1,......, 2^d$ are $D$ indices. 
Now, using equation \ref{eq:haarintegral},
\begin{equation}
\begin{aligned}
    \int dU \text{Tr}(\rho^2_{AC}) &= \frac{1}{d^2}\left( \frac{1}{d^2-1}(d_Ad_B^2d_Cd_D^2 + d_A^2d_Bd_C^2d_D) - \frac{1}{d(d^2-1)}(d_Ad_B^2d_C^2d_D + d_A^2d_Bd_Cd_D^2) \right)       
\end{aligned}
\end{equation}
We can approximate above equation as:
\begin{equation}
    \int dU \text{Tr}(\rho^2_{AC}) = d_A^{-1}d_C^{-1} + d_B^{-1}d_D^{-1} - d^{-1}d_A^{-1}d_D^{-1} - d^{-1}d_B^{-1}d_C^{-1} 
\end{equation}
From these quantities, it is possible to obtain the expression for R\'enyi enropies, which allow us to obtain bound for other information theoretic quantities.

\section{Information Scrambling}
\label{sec:scrambling}
\begin{figure}
    \centering
    \includegraphics[scale=1.0]{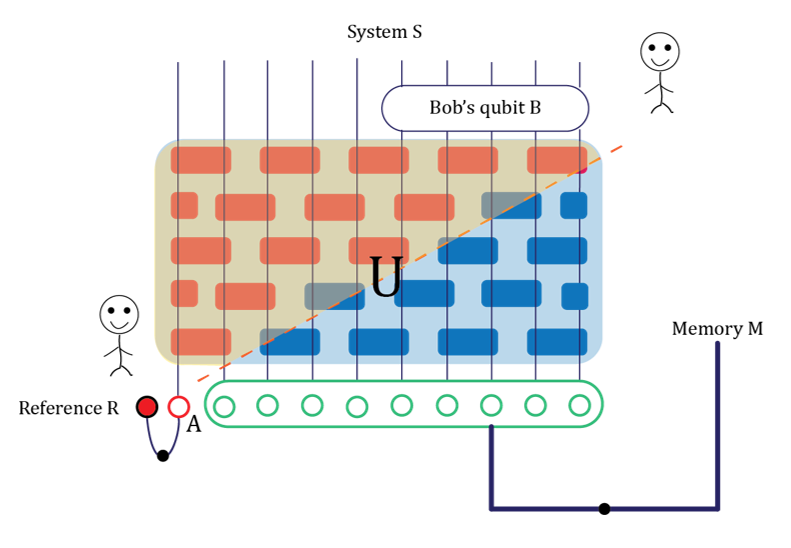}
    \caption{Alice and Bob try to communicate via a scrambling unitary $U$ acting on N qubits. Alice has full control of the first qubit $A$, which is entangled with reference system $R$. The rest of the qubits, marked green, can be in a mixed state  $\rho$ purified by an external system called Memory. This pure state is denoted by $\ket{\sqrt{\rho}}$.  The scrambling unitary acts only on Alice's qubit and the systems qubit, which is marked green. After the unitary evolution, Bob then has access to the set of the qubits of the system. If the scrambling unitary has a local structure, it is also possible to have an information light cone indicated by the yellow cone. The initial state of the system, including the reference, system, and memory, is:
     $\ket{\boldsymbol{\psi}} = \frac{1}{\sqrt{2}}(\ket{0}_R\ket{0}_{q_A} +\ket{1}_R\ket{1}_{q_A}) \ket{\sqrt{\rho}}$
Since the time evolution unitary doesn't act on reference and memory, the total dynamics is then given by:
$\ket{\boldsymbol{\psi}(t)}  = \frac{1}{\sqrt{2}} \left(\ket{0}_R U_S \otimes I_\text{MEM} \ket{0,\sqrt{\rho}}  +  \ket{1}_R U_S \otimes I_\text{MEM} \ket{1,\sqrt{\rho}} \right) $. }
    \label{fig:informationScrambling2}
\end{figure}

The phenomenon of "scrambling" of quantum information is a fundamental concept closely related to many-body physics, quantum chaos, complexity theory, and black holes \cite{r10,r11}. This phenomenon can be seen as a form of "thermalization of the quantum information" and is crucial in understanding information spreading in quantum systems.
As illustrated in Figure \ref{fig:informationScrambling2}, at time $t=0$, the reference $R$ is only entangled with the Alice qubit $q_A$, such that $S(R) = S(q_A) = 1$, $S(Rq_A) = 0$, and $I(R:q_A) = 2$. As time progresses, the quantum information from Alice's qubit becomes more and more scrambled, depending on the system's dynamics. This is because the initially localized entanglement of Alice's qubit with the reference $R$ begins to expand across the system. In one scenario, the dynamics can be such that the information is carried coherently, for example, through a circuit using only SWAP gates. The information is highly localized in this case, and the unitary cannot be considered scrambling. Conversely, in strongly interacting systems, the entanglement spreads in a complex fashion such that the reference $R$ becomes entangled with a large collection of qubits. In this case, the information is considered to be scrambled. Because quantum mechanics is unitary, information, however, is not lost globally. 

In the literature of quantum chaos \cite{larkin, Maldacena_2016,Hashimoto_2017}, one usually defines the information scrambling in terms of the spreading of operators. For this, we consider a unitary map $U_{AB}: A \otimes B \xrightarrow{} C \otimes D$, and let $O_i$ be hermitian operators supported on region $i$, where $i \in {A,B,C,D}$ and define $O_i(t) = U^\dagger O(0) U$. 
We then define the $U_{AB}$ as chaotic scrambling \cite{yoshida2017efficient} iff it satisfies:
\begin{equation}
\label{eq:chaoticscrambling}
  \langle O_A(0) O_B(t) O_C(0) O_D(t) \rangle\simeq \langle O_AO_C \rangle \langle O_B \rangle \langle O_D \rangle + \langle O_BO_D \rangle \langle O_A \rangle \langle O_C \rangle - \langle O_A \rangle \langle O_B \rangle \langle O_C \rangle \langle O_D \rangle  
\end{equation}
 where $\simeq$ means up to an order $d^{-2}$. Studying operator spreading in arbitrary systems might be too complicated.
 
 Instead, we follow a trick due to Wigner \cite{Wigner1957} and pick a unitary $U_{AB}$ at random from a Haar measure. Indeed, we expect the late-time values of entropy and mutual information for any scrambling unitary $U(t)$ to match those of Haar random unitaries. To separate from chaotic scrambling, we define a Haar scrambled quantum state as the state obtained by applying a random unitary $U$ chosen from group-invariant Haar measure, $\ket{\psi (U)} = U \ket{\psi_0}$. The Haar measure has several interesting mathematical properties that make it possible to obtain analytical results for different information theoretic quantities \cite{mele2023introduction}.
 One important physical characteristic of Haar scrambling is that a randomly chosen pure state in $\mathcal{H}_{AB}$ is likely close to a maximally entangled state if $\frac{|A|}{|B|} << 1$. This property is called Page scrambling, named after Page, who first used it to explain black hole physics \cite{Page_1993}. Mathematically, for any bipartite Hilbert space $\mathcal{H}_A \otimes \mathcal{H}_B$, we have: 
\begin{equation}
    \int dU || \rho_A(U) - \frac{I_A}{|A|} ||_1 \leq \sqrt{\frac{|A|^2 - 1 }{|A||B| +1 }}
\end{equation}
and when $|B|$ is significantly larger than $|A|$, the typical deviation of $\rho_A$ from maximally mixed state is extremely small. For example, if $B$ has 10 more qubits than $A$, then the typical deviation from maximal entanglement is bounded by $2^{-5}$. As an example, let us consider a case when $A$ is a single qubit and let $p$ be the size of region $C$. Using the analytical results, we can upper-bound the quantum mutual information between $R$ and $C$ as
\begin{equation}
\label{eq:R'enyiMutualInformation}
I(R:C) \leq 1 + \log_2 \left( 2 - \frac{3(1 - 2 ^{2p -2 N})}{2 + (2^{-s} - 2^{-N + 1})4^{p- N/2}} \right),
\end{equation}
where $s$ is the second R\'enyi entropy of system $B$. 

\subsection{Quantum error correction from scrambling}
\label{sec:RampSecret}
Let the seed information $S$, described by a quantum state $\rho_S$ in a Hilbert space $\mathcal{H}_S$ of dimension $K$, be purified by a reference system $R$ with Hilbert space $\mathcal{H}_R$. The resulting purification is denoted by $  |S R\rangle \in \mathcal{H}_S \otimes \mathcal{H}_R$.

We denote by $\ket{RX_1 \dots X_n}$ the state of the quantum system $RX_1 \dots X_n$ after applying a completely positive isometric encoding map $V$ to $S$ and identity to $R$. The $n$ quantum systems $X_1, \dots , X_n$, each of level $q$, is partitioned into two arbitrary disjoint blocks $X_I$ and $X_J$, of size $|I|=n-d+1$ and $|J|=d-1$, respectively. A quantum erasure error correcting code of distance $d$ allows the reconstruction of the seed information $\rho_S$ even after erasing block $X_J$. This is possible if the following two information-theoretic requirements are satisfied \cite{r9,Grassl_2022}:
\begin{itemize}
    \item Recoverability: Recovery of seed information: $I(R:X_I) = I(R:S)$
    \item Secrecy: No information by erasures: $I(R:X_J)=0$
\end{itemize}
This version of quantum error correction is focused on erasure errors, while a standard error correction allows for the correction of arbitrary errors \cite{NielsenChuang2000}. However, it is much easier to use information theoretic formalism for erasure errors than standard errors, and it will be sufficient for us. In fact, this scheme is equivalent to another cryptographic primitive known as quantum secret sharing schemes \cite{r7, Hillery_1999, r8, r9}. Here, an arbitrary quantum secret is divided into $n$ parties such that any $k$ or more parties can perfectly reconstruct the secret, while any $k-1$ or fewer parties have no information at all about the secret. This also means that the reduced density matrix of these $k-1$ parties is independent of the secret and denoted by the $((k,n))$ threshold scheme.

For example, consider this $((2,3))$ quantum threshold scheme. Here a dealer share an arbitrary qutrit among three parties $V_{2,3}:\mathcal{C}^3 \xrightarrow{} \mathcal{C}^3 \otimes \mathcal{C}^3 \otimes \mathcal{C}^3$ defined as:
\begin{equation}
\begin{split}
        V_{2,3}(\alpha \ket{0} + \beta \ket{1} + \gamma \ket{2}) = \frac{1}{\sqrt{3}}(&\alpha (\ket{000} + \ket{111} + \ket{222} ) + \\
        & \beta ( \ket{012} + \ket{120} + \ket{201}) + \\
        & \gamma (\ket{021} + \ket{102} + \ket{210}))
\end{split}
\end{equation}
Indeed, $V_{2,3}$ is an isometry. Each resulting qutrit is taken as a share. From a single share, no information can be obtained about the secret as it is totally mixed, and, therefore, it is independent of the secret.
\begin{equation}
\rho_1 = \frac{1}{3}(\ket{0}\bra{0} + \ket{1}\bra{1} + \ket{2}\bra{2})
\end{equation}
However, the secret can be perfectly reconstructed from any two of the three shares. For example, if we have two first shares, then the secret can be reconstructed as:
\begin{equation}
   (\alpha \ket{0} + \beta \ket{1} + \gamma \ket{2}) (\ket{00} + \ket{12} + \ket{21})
\end{equation}
So, the first qutrit now contains the secret. Here, the reconstruction procedure can be done by adding the value of the first share to the second (modulo three) and then adding the value of the second share to the first. For the other cases, we can follow a similar procedure because of its symmetric nature.

Unfortunately, a scrambling unitary rarely satisfies the perfect threshold secret-sharing scheme. Instead, we will now show that it satisfies a ramp secret-sharing scheme where partial information about the secret is allowed to leak to a set of participants who cannot fully reconstruct the secret. In particular, there are parameters $1 \leq b < g \leq n$ such that set of parties of size at least $g$ should be able to reconstruct the secret,  while sets of parties of size at most $b$ cannot learn anything about the secret. Furthermore, we make no requirement on the parties of size $g-b$. $g$ and $b$ stand for good and bad, respectively. Such a scheme is denoted by $((b,g))$.

\begin{theorem}
\label{theorem:purestate}
A scrambling unitary $U$ drawn from the Haar measure satisfies $\left( \left( \frac{N}{2}- \epsilon, \frac{N}{2}+\epsilon \right)  \right)$ approximate quantum ramp secret sharing scheme when the quantum secret is encoded into pure state players.

\begin{proof}
 Let $P(l)$ be the arbitrary region in the system  of size $l$ after the Haar scrambling unitary is applied, and $I(P(l))$ be the mutual information between the original secret, of size one qubit, and the region $l$ which we know from  Eq: \ref{eq:R'enyiMutualInformation} satisfies the inequality, 
\begin{equation*}
    I(P(l)) \leq  1 + \log_2 \left(  2 - \frac{3(1 - 2 ^{2l -2 N})}{2 + 4^{l- N/2}}        \right)
\end{equation*}
For the arbitrary subsystem of size $l = \frac{N}{2} - \epsilon$,
\begin{equation*}
\begin{aligned}
 I(P(l))   & \leq 1 + \log_2 \left(  2 - \frac{3(1-2^{ -2\epsilon -N})}{2+4^{-\epsilon}}    \right)  \\
  &\leq 1 + \log_2 \left( 2 - \frac{3}{2 + 4^{-\epsilon}} \right)  \\
  &\leq 2 - \frac{3}{\ln 2}4^\epsilon
\end{aligned}    
\end{equation*}
where we used the relation $\log_2(1+ x) \leq \frac{x}{\ln2}$ and taken the large $N$ limit. If we pick $\epsilon = \log \left( \sqrt{\frac{\ln2}{3}\left( \frac{2}{3} - \mathcal{O}(\gamma) \right)} \right)$, then we obtain the relaxed secrecy requirement:
\begin{equation}
    I(P(l)) \leq  \mathcal{O}(\gamma)
\end{equation}
The information about the original quantum secret should be preserved under unitary transformation $U$. For the pure state scheme, 
\begin{equation}
\label{eq:boundForPureCase}
    I(P(l)) + I(R:P(N-l)) = I(R:S)
\end{equation}
From this, we also get the relaxed recoverability requirement:
\begin{equation}
    I(R:N-l) \geq I(R:S)-   \mathcal{O}(\gamma)
\end{equation}
Therefore, this corresponds to the $\left( \left( b, g \right)  \right)$ approximate quantum ramp secret sharing scheme with parameters:
\begin{equation}
       g = \frac{N}{2}+\epsilon , b = \frac{N}{2} - \epsilon
\end{equation}
\end{proof}
\end{theorem}
This condition can be improved if we instead consider the initial secret as a mixed state. This is because mixed-state encoding can always be purified by adding an extra share.

\begin{theorem}
A scrambling unitary $U$ drawn from the Haar measure satisfies  $ \left( \left( \frac{N+s}{2}- \epsilon, \frac{N+s}{2}+\epsilon \right) \right)$  approximate quantum ramp secret sharing scheme, where $s$ is extra players necessary to purify $N$ plazers. 

\begin{proof}
We start with a $((b,g)$ pure approximate quantum ramp scheme with the total $N'$ players.
From Theorem \ref{theorem:purestate}, the ramp parameters for the pure case are:
\begin{equation}
\begin{aligned}
    b &= \frac{N'}{2} + \epsilon \\
    g &= \frac{N'}{2} - \epsilon
\end{aligned}
\end{equation}
Suppose we discard  $s$ extra players. However, the parameters $b$ and $g$ don't change as long as $g \leq N$. 
Therefore, we get a $ \left( \left( \frac{N+s}{2}- \epsilon, \frac{N+s}{2}+\epsilon \right) \right)$ for a mixed state ramp scheme.
\end{proof}
\end{theorem}
\end{appendices}

\acknowledgments
The research is part of the Munich Quantum Valley, which is supported by the Bavarian state government with funds from the Hightech Agenda Bayern Plus. This research is also part of the Abdus Salam International Centre for Theoretical Physics (ICTP) program: Physics Without Frontiers (PWF) and we acknowledge support from the PWF program of the ICTP, Italy. I also acknowledge the online resources—particularly recorded lectures, conference presentations, and virtual seminars—which have greatly aided my research on the black hole information paradox.

\bibliographystyle{JHEP}
\bibliography{ref1.bib}
\end{document}